\documentclass{rspublic}
 
\usepackage{epsf}
\usepackage{longtable}
\usepackage{natbib}
\usepackage{epstopdf}
\usepackage{graphicx}

\begin{document} 
 
\title[MHD Waves and Coronal Seismology]{MHD Waves and Coronal Seismology: an overview of recent results} 
 
\author[I.~De Moortel, V.M.~Nakariakov]{Ineke De Moortel$^1$, Valery M. Nakariakov$^{2,3}$} 
 
\affiliation{$^1$School of Mathematics \& Statistics, University of St Andrews, North Haugh, St Andrews KY16 9SS, UK\\$^2$Centre for Fusion, Space and Astrophysics, Department of Physics, University of Warwick, Coventry CV4 7AL, UK\\ 
$^3$ Central Astronomical Observatory of the Russian Academy of Sciences at Pulkovo, 196140 St Petersburg, Russia } 
 
\maketitle 
 
\begin{abstract}{MHD Waves, Coronal Seismology, Corona} 
Recent observations have revealed that MHD waves and oscillations are ubiquitous in the solar atmosphere, with a wide range of periods.  We give a brief review of some aspects of MHD waves and coronal seismology which have recently been the focus of intense debate or are newly emerging. In particular, we focus on four topics: (i) the current controversy surrounding propagating intensity perturbations along coronal loops, (ii) the interpretation of propagating transverse loop oscillations, (iii) the ongoing search for coronal (torsional) Alfv\'en waves and (iv) the rapidly developing topic of quasi-periodic pulsations (QPP) in solar flares. 
\end{abstract} 
 
\section{Introduction} 

The study of MHD waves has two major applications within solar physics, namely, coronal (or magneto) seismology (\citealt{Uchida70,Roberts84}) and the role of MHD waves in coronal heating. Both topics have been the subject of years of study. As MHD waves can carry magnetic energy over large distances, it was historically thought that they could play a major role in the heating of the solar atmosphere, especially in open field regions. However, a lack of actual observations in the solar atmosphere meant that, for a long time, studies of MHD waves were mainly theoretical. This situation has been changed dramatically over the last two decades with the advent of both imaging and spectroscopic instruments with high spatial and temporal resolution. Over the years, the observations have gradually revealed that waves and oscillations, which fall within the MHD spectrum, are present in most, if not all, coronal structures and that these waves can potentially provide a considerable part of the energy needed to heat the (quiet) solar corona and drive the solar wind. These observations have led to a rapid development of coronal seismology as well as renewed interest in the role of waves and oscillations in the heating of both open \emph{and} closed field regions. Examples of both standing and propagating, slow and fast mode oscillations in coronal loops have been identified, as well as large-scale, global coronal perturbations.
    
To give a comprehensive overview of MHD waves, their role in coronal heating and the booming field of coronal seismology is impossible within this limited review. For comprehensive (recent) reviews on the subjects of coronal heating, we refer the interested reader to e.g.~\cite{Walsh03,Erdelyi04,Ofman2005,IDM08} or \cite{Taroyan2009}. Reviews of MHD waves and coronal seismology can be found in \cite{Nakariakov05, IDM05, Erdelyi2006} or \cite{Banerjee2007}, to name but a few.  During 2009 and 2011, two \emph{volumes} of Space Science Reviews were dedicated to detailed descriptions of various aspects of MHD wave propagation and coronal seismology. In this short paper, we will focus on topics which are presently most relevant, either because they are newly emerging or currently generating substantial debate. We will try to provide the reader with some context and outline the broad lines of the current thinking, rather than reporting details of individual studies. In particular, we will not include standing kink mode and standing slow mode oscillations of coronal loops (see \citealt{Ruderman2009,Terradas2009, Wang2011}), global coronal oscillations (see \citealt{WillsDavey2009,Gallagher2011}) or oscillations in prominences (see \citealt{Oliver2009,Tripathi2009}).

 In Section 2, we focus on quasi-periodic perturbations propagating along coronal loops (the current flows vs waves debate), in Section 3 we highlight the recent results on Alfv\'enic, propagating waves and in Section 4, we  discuss the current search for torsional (coronal) Alfv\'en waves. In Section 5 we summarise the currently emerging topic of QPP's. In Section 6 we highlight some recent developments in coronal seismology and conclude this review with a series of open questions.

 \section{Propagating Periodic Disturbances} 

Observations of disturbances traveling along coronal structures have been reported by a number of authors since 1999. However, recently the interpretation of these events as propagating slow magneto-acoustic waves has come under renewed scrutiny. For a more comprehensive description (especially of the earlier results), we refer the interested reader to the detailed reviews by \cite{IDM06,IDM09} and \cite{Banerjee2010}. 

\begin{figure}[ht]
\centering
%\scalebox{.25}{\includegraphics{DePontieu2010_fig1_cropa.jpg}}
%\scalebox{.4}{\includegraphics{fig2i_nbk.jpg}}
%\scalebox{.25}{\includegraphics{DePontieu2010_fig1_cropb.jpg}}
%\scalebox{.4}{\includegraphics{fig2v_nbk.jpg}}
\scalebox{.25}{\includegraphics{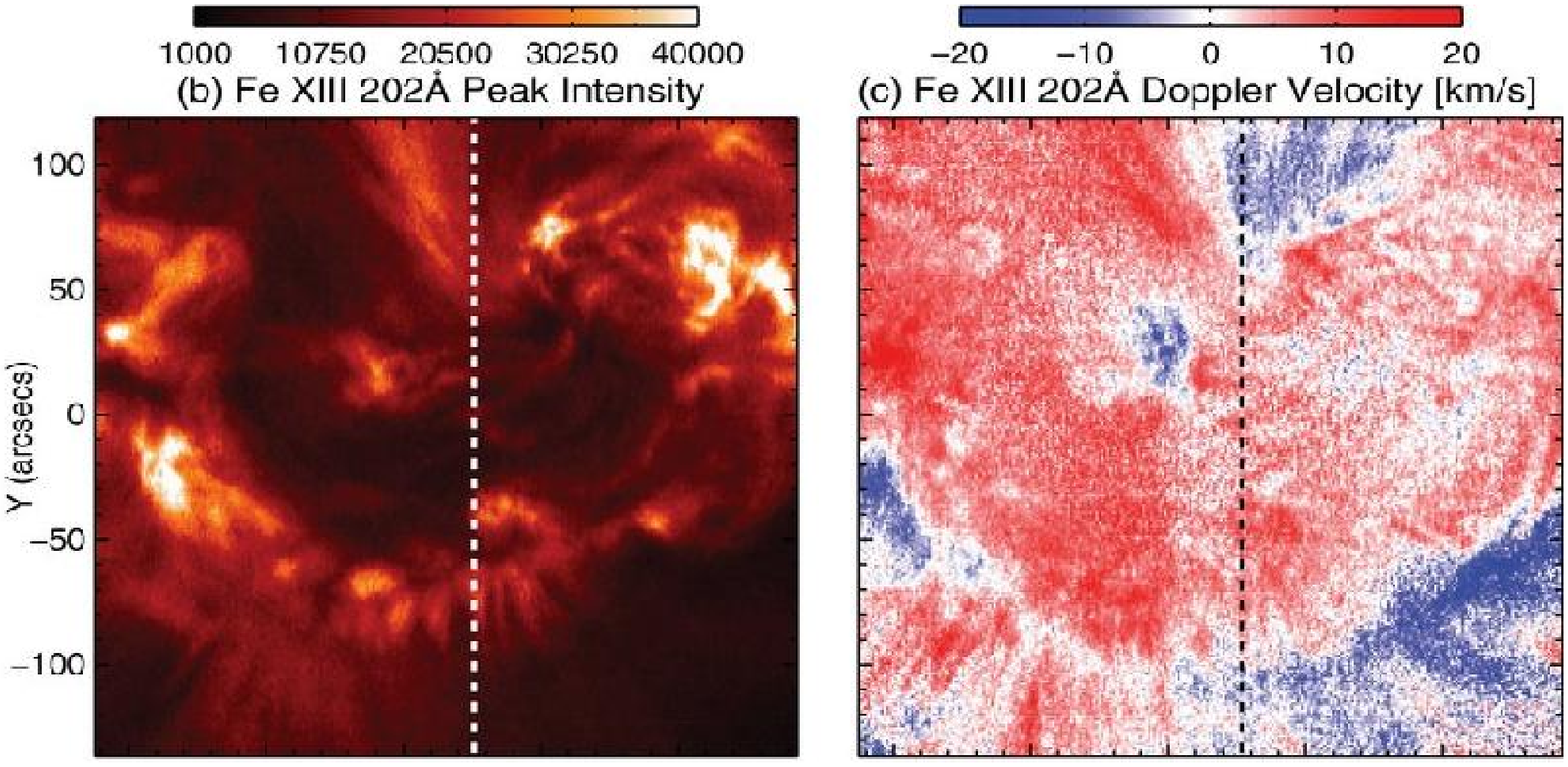}}
\scalebox{.4}{\includegraphics{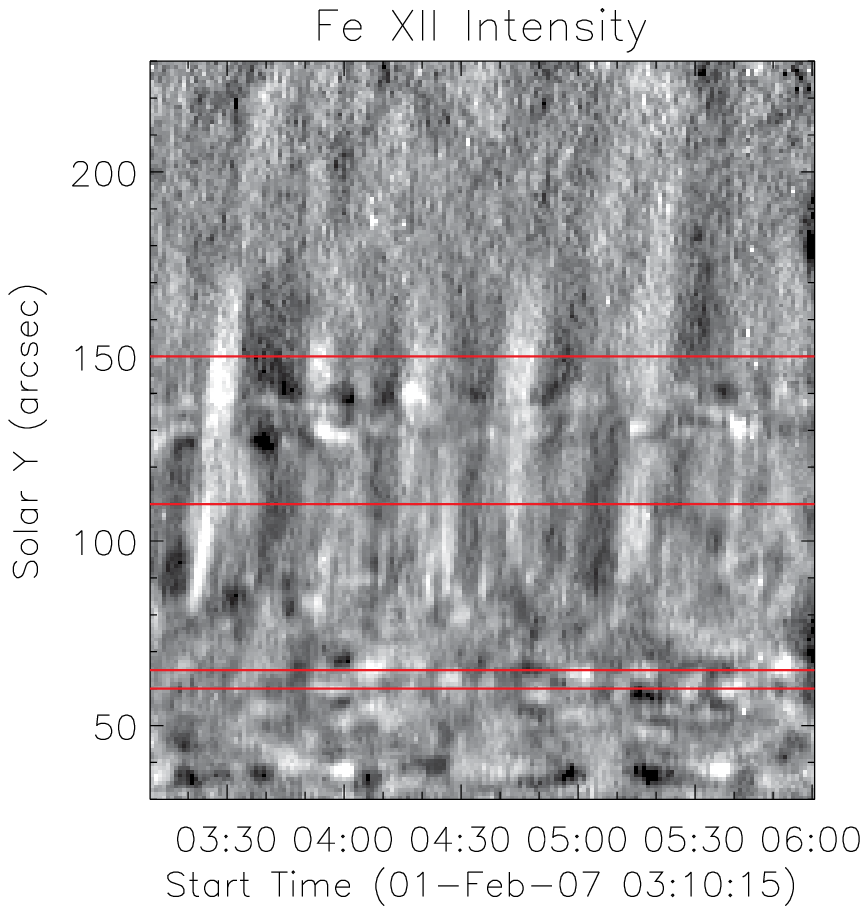}}
\scalebox{.25}{\includegraphics{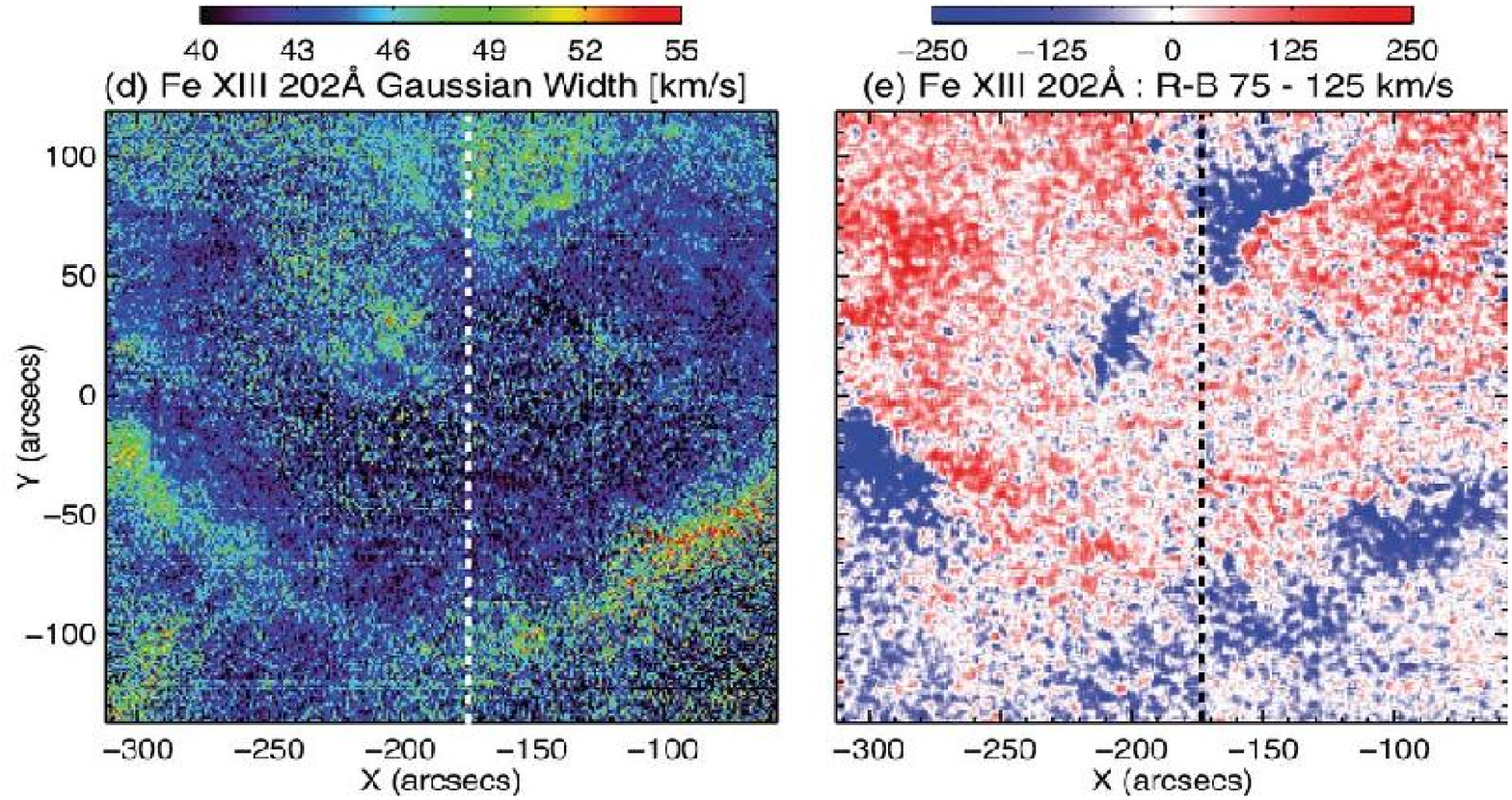}}
\scalebox{.4}{\includegraphics{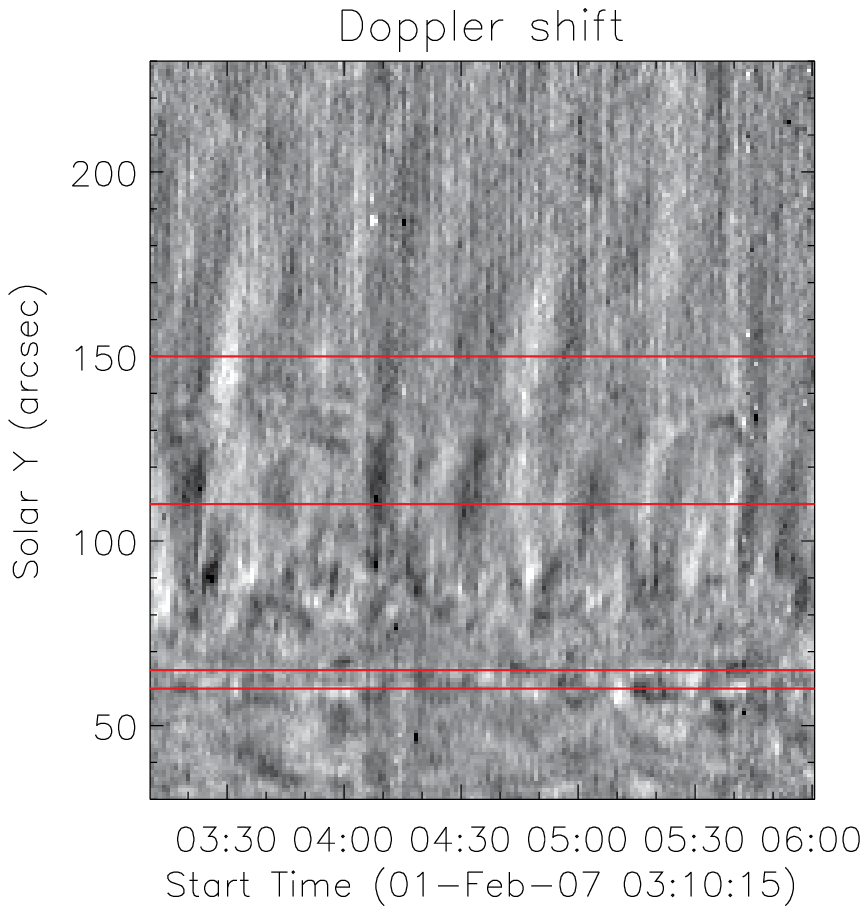}}
\caption{(Left) The region analysed by both \cite{DePontieu2010} and \cite{Wang09b}. Inferred single Gaussian fit parameters to the EIS spectroheliogram showing the peak intensity (panel (b)), (relative) Doppler velocity (panel (c)), Gaussian width (panel (d)), and the results of the 75-125 km/s R-B analysis in Fe {\sc xiii} 202 \AA\ (panel (e)) from \cite{DePontieu2010}. On each of the panels the pointing of the time series is also shown (vertical dashed line). (Right) The upwardly-propagating waves in coronal loops observed by Hinode/EIS, reported by \cite{Wang09b}. (top) Time series of relative intensity along the slit in the Fe {\sc xii} 195.12 \AA\ line. (bottom) Time series of Doppler shift. Here the white color indicates the blueshift and the black color indicates the redshift.}
\label{fig:DePontieu2010-fig1}
\end{figure}

Reporting on the TRACE first results, \cite{Schrijver99} describe ``upward motions in the fans of 1MK loops in the outer envelope of the active-region corona.'' These authors argue that the low propagation speed ($\sim$40 km/s) makes an interpretation in terms of (MHD) wave modes implausible but requires the presence of flows. A similar conclusion was reached by \cite{Winebarger02}. Following these initial reports, propagating disturbances were reported by a number of authors in both large coronal loops and (polar) plumes (see e.g.~\citealt{Berghmans1999,IDM00,Ofman97,Banerjee2000,OShea2006}). Typically, the perturbations have amplitudes of a few percent of the background intensity, disappear below the noise level within a (coronal) gravitational scale height, display periods in the range of 2-10 minutes and velocities of the order of 100 km/s (see \cite{McEwan06}). It was the combination of these properties, but especially the apparent match with the local sound speed, that lead to the interpretation as propagating slow magneto-acoustic waves. Theoretical modelling (\citealt{Nakariakov00,Tsiklauri01,Ofman00,IDM03,IDM04,IDM04-2}) confirmed that the EUV imaging observations could be interpreted in terms of propagating slow magneto-acoustic waves, with thermal conduction proposed as the main damping mechanism and the quasi-periodic nature of the waves attributed to the leakage of p-modes from the solar interior into the corona (\citealt{DePontieu05,DePontieu06,IDM07,Erdelyi07,Malins07}). As this model could account for the major observational properties, it quickly became established as the prevalent interpretation.

Recently, the EUV imaging observations have been complimented by spectroscopic observations from Hinode/EIS and the picture has become considerably more complicated. The spectroscopic data show similar evidence of low amplitude, quasi-periodic oscillations not only in intensity but also in the Doppler velocity. However, the interpretation of the observed perturbations has (again) come into question. Some authors still favour the propagating slow magneto acoustic wave interpretation (\citealt{Marsh2009,Banerjee2009b,Wang09a,Wang09b,Kitagawa2010,Mariska2010,KrishnaPrasad2011,Marsh2011,Wangetal11}) but other authors have interpreted quasi-periodic disturbances with very similar properties as (quasi-periodic) upflows (\citealt{Sakao07,Doschek07,Doschek08,2008A&A...481L..49D,Harra08,Hara08,DePontieu09,McIntosh2009a,McIntosh2009b,He2010,Guo2010,McIntosh2010,Peter2010,Bryans2010,Tian2011,Ugarte-Urra2011,Warren2011}). Several papers have reported on quasi-periodically occurring enhancements in the blue-wing of the spectral line profiles that are co-located with the propagating disturbances, with motions of the same order of magnitude. These enhancements are revealed when assessing the asymmetry, or fitting the lines with a double rather than a single Gaussian model. The need for a careful fitting of spectral lines using a double Gaussian model has recently been highlighted by several authors (e.g.~\cite{Peter2010}). Detailed analysis of spectroscopic line profiles by \cite{DePontieu2010} and \cite{Tian2011} shows that fitting line profiles which exhibit a quasi-periodically occurring excess in the blue wing with a single Gaussian will mimic the properties of the disturbances observed in the imaging data. However, an alternative interpretation of the quasi-periodically varying line profiles, again in terms of propagating slow waves,  was put forward by \cite{Verwichte2010} and \cite{Wangetal11}, by suggesting that the (varying) double-Gaussian fit could consist of an oscillating dominant (core) component and an additional small, stationary blue-wing component. 

\begin{figure}[t]
\centering
%\scalebox{2.15}{\includegraphics{DePontieu2010_fig5.jpg}}
\scalebox{2.15}{\includegraphics{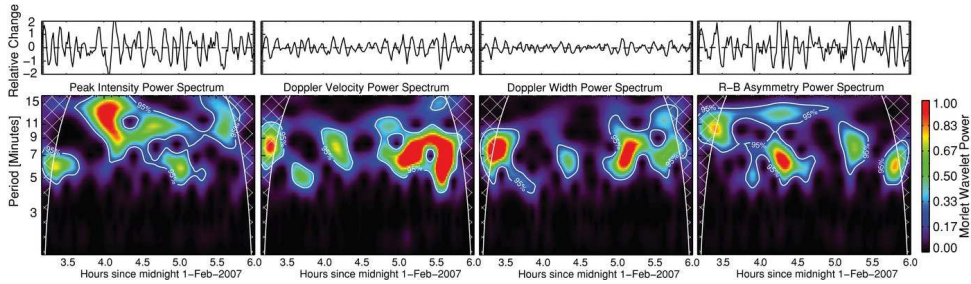}}
\caption{Wavelet power spectra for the Fe {\sc xiii} 202 \AA\ time series of \cite{DePontieu2010}. The solid white contours denote regions of the wavelet power spectrum of 95\% significance and the cross-hatched region encloses the cone-of-influence for the power spectrum.}
\label{fig:DePontieu2010-fig5}
\end{figure}

Distinguishing between waves and flows is less straightforward than one might expect. It has become apparent that any distinguishing characteristics identified so far require extensive analysis of the spectroscopic and imaging data, complicated further by the low signal-to-noise ratio of these small amplitude perturbations. Therefore, it is not surprising that, at present, different authors reach different conclusions, even on the same datasets, as the observational signatures are difficult to disentangle (\citealt{Verwichte2010,Peter2010,DePontieu2010,Tian2011}). Compare for example \cite{Wang09b} and \cite{DePontieu2010}, who both analyse the region shown in Fig.~\ref{fig:DePontieu2010-fig1}(left) in great detail, studying the intensity, Doppler velocity, line widths and line asymmetry in a variety of spectral lines. Fig.~\ref{fig:DePontieu2010-fig1}(right) (taken from \cite{Wang09b}) shows the running difference plots, which are typically used to identify propagating disturbances in imaging data. Analysing spectroscopic EIS data, \cite{Wang09b}) find that the oscillations in intensity and Doppler shift are (approximately) in phase, which these authors interpret as evidence of propagating slow magneto acoustic waves. However performing further analysis on the same dataset, \cite{DePontieu2010} show that significant, in-phase, oscillations are found not only in the intensity and Doppler velocity, but also in the line widths and line asymmetries (see Fig.~\ref{fig:DePontieu2010-fig5}). These latter authors use modelling to show that such in-phase behaviour of oscillations in intensity, Doppler velocity, line widths and line asymmetries can be explained in terms of quasi-periodic upflows.   

Regardless of the differing interpretations of the exact nature of these propagating disturbances, their quasi-periodicity has been clearly established. The periods of the observed perturbations are of the order of a few to ten minutes, a timescale which links them to the solar surface perturbations (p-modes) or convective buffeting (\citealt{Fedun2009,Hansteen2010,Fedun2011}). It is quite likely that the buffeting and tangling of the magnetic field by these solar (sub)surface perturbations will generate disturbances travelling along the magnetic field. However, a definitive identification of the driver has not been established. Also, it is unclear what the nature of these perturbations is (flows, waves, both or something entirely different?) and what happens to them as they travel through the solar atmosphere, although the similarity in properties hints at a close relation with chromospheric ``Type II'' spicules (\citealt{DePontieu09,Rouppe2009}). These propagating disturbances have been linked with the mass cycle of the solar corona and the solar wind (e.g.~\citealt{McIntosh2009a,McIntosh2010,DePontieu2011}) and it has been realised that despite their small amplitude, their omnipresence in the solar atmosphere could make them a significant player in the coronal energy budget. Very recent work by \cite{McIntosh2012} reports on slower, counterstreaming downflows ($\sim$10 km/s) in cooler lines, which the authors interpret as the return-flow of coronal material (i.e.~the end of the mass cycle). In sunspots, recent numerical simulations of acoustic 3-min oscillations demonstrated that sunspot umbrae act as a non-ideal resonator, and that the leaky part of the oscillations naturally develops into the propagating waves in the corona {\citealt{Botha2011}). Finally, we point out that the damping of the slow waves by thermal conduction also appears less robust than previously thought. Indeed, using an interpretation in terms of slow magneto-acoustic waves, \cite{Marsh2011} point out that for longer periods, thermal conduction cannot account for the observed rapid damping of the perturbations.

\section{Propagating Transverse Loops Oscillations} 
 
Standing, transverse loop oscillations were one of the first examples of coronal loop oscillations to be observed and studied in great detail (see e.g. \citealt{Nakariakov1999,Aschwanden2002,Schrijver2002}). These oscillations, often observed to be generated by a nearby impulsive event, have generally been interpreted as (fast) kink mode oscillations (see \citealt{Ruderman2009,Terradas2009,Goossens2009} for a detailed review). Recently, there have been several reports in the literature of the observational detection of transverse perturbations, {\it propagating} along the magnetic field and localised in the transverse direction \citep{2007Sci...318.1580C, 2007Sci...318.1577O, 2007Sci...317.1192T}.  These propagating displacements were detected with imaging instruments Hinode/SOT and Hinode/XRT in both cool and hot coronal structures, such as a prominence fibril \citep{2007Sci...318.1577O} and soft X-ray jets \citep{2007Sci...318.1580C}, as well as in (chromospheric) spicules by \cite{DePontieu07} and \cite{He2009a,He2009b} (see also \citealt{Zaqarashvili2009}). Similar perturbations were seen in large, off-limb, coronal loops by the Coronal Multi-Channel Polarimeter instrument (COMP) as periodic Doppler shifts \citep{2007Sci...317.1192T} propagating along coronal magnetic field lines. As we are focusing this review on oscillations in coronal loops, we will only provide some detail of the COMP studies but the other observations have revealed largely similar properties.

The COMP observations \citep{2007Sci...317.1192T,Tomczyk2009} indicate that the propagating transverse perturbations are, both spatially and temporally, ubiquitous in the solar corona. Fig.~\ref{fig:Tomczyk} shows a snapshot of the intensity, 3.5 mHz-filtered Doppler velocity and power ratio (i.e. outward power/inward power) observed by COMP.  The waves were detected in time series of Doppler images but did not cause significant perturbations in intensity nor noticeable loop displacements, explaining why they have not been seen earlier by imaging instruments.  The observed periods are of the order of several minutes, with a relatively broad power spectrum peaking at 5 minutes. This focus on 5 minute periods hints that the (footpoint) driving is likely to be related to the solar surface perturbations (p-modes). The propagation speed was estimated by \cite{Tomczyk2009} to be of the order of 600 km/s. This is  significantly higher than the local sound speed and places the observed propagation speeds in the range of the expected Alfv\'en speed. The observed speed is roughly constant in time, indicating that the structure supporting the oscillations (the wave guide) is relatively stable in time. The correlation analysis of \cite{2007Sci...317.1192T}, showed that the waves are indeed extended along the magnetic field, as the correlation length substantially exceeds the correlation width. Finally, there is a clear discrepancy between outward and inward power, with significant inward power only observed along shorter coronal loops (with footpoint separation $<$ 300 Mm). This indicates that the observed propagating perturbations are subject to considerable in-situ damping.

\begin{figure}[t]
\centering
%\scalebox{.45}{\includegraphics{InekeFigure.pdf}}
\scalebox{.45}{\includegraphics{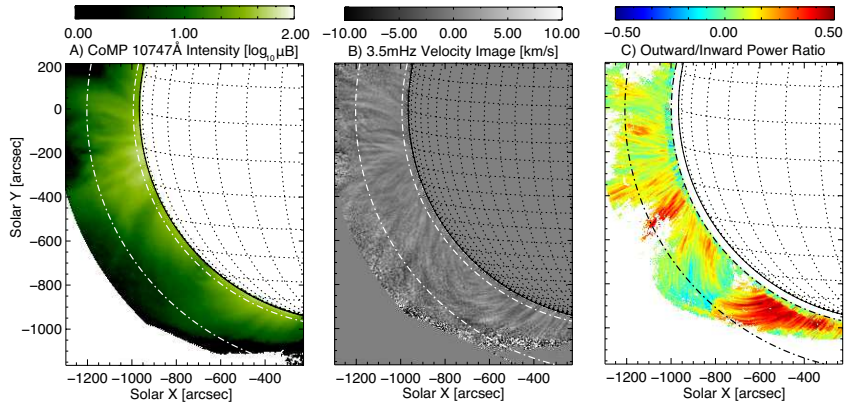}}
\caption{Snapshot of CoMP observations of time-averaged intensity (left),  3.5-mHz filtered Doppler velocity (middle) and power ratio map (right) taken from \citealt{2007Sci...317.1192T,Tomczyk2009}}
\label{fig:Tomczyk}
\end{figure}

With speeds in the regions of the (estimated) local Alfv\'en speed, no evidence of significant intensity perturbations (and hence largely incompressible) and the apparent restoring force the magnetic tension, these observed propagating perturbations have characteristics which are Alfv\'enic in nature. Hence, they were  originally interpreted as propagating (shear) Alfv\'en waves. Subsequent theoretical studies suggested a different interpretation of the observed displacements as propagating kink modes (\citealt{ErdelyiFedun07,2008ApJ...676L..73V, 2008A&A...491L...9V, 2009A&A...498L..29V}). Indeed, the kink mode is locally a fast magetoacoustic wave, propagating obliquely to the magnetic field and guided along the field by a field-aligned plasma structure (a waveguide) by reflection or refraction (see e.g.~\citealt{Nakariakov1996}). In the observations, the perturbed magnetic flux tube was displaced in the transverse direction as a whole, and the transverse size of the perturbation was at least an order of magnitude shorter than the wavelength along the field. In a 2D numerical study, \cite{2008A&A...491L...9V} pointed out the necessity for transverse structuring as without such a waveguiding field-aligned plasma non-uniformity, the perturbations would propagate not along the field, as observed, but across it. This is connected with the competition of two restoring forces, the magnetic tension force and the gradient of the total pressure. 

\begin{figure}[t]
\centering
%\scalebox{.23}{\includegraphics{modecoupling.jpg}}
%\scalebox{.18}{\includegraphics{f5.jpg}}
\scalebox{.23}{\includegraphics{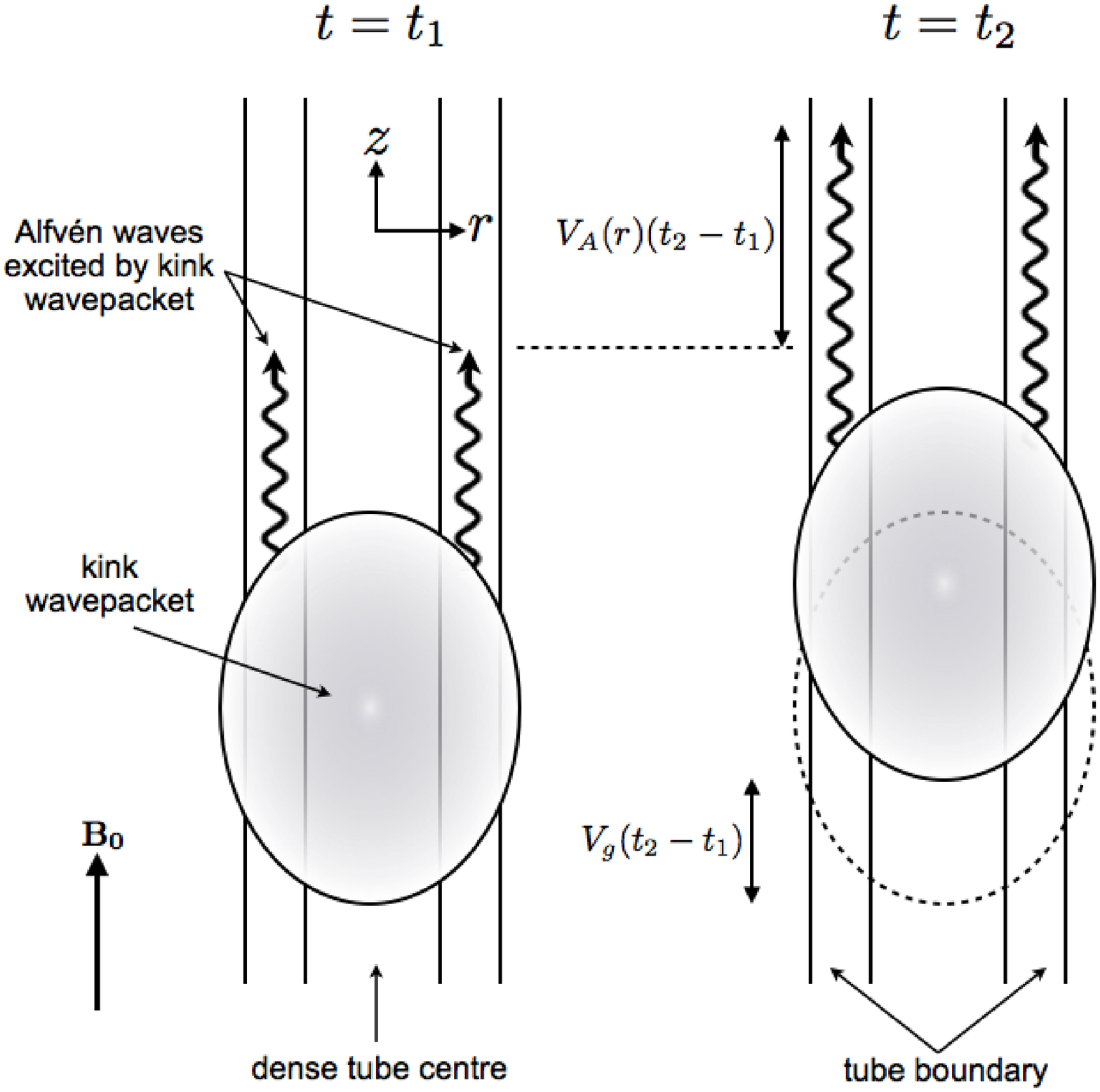}}
\scalebox{.18}{\includegraphics{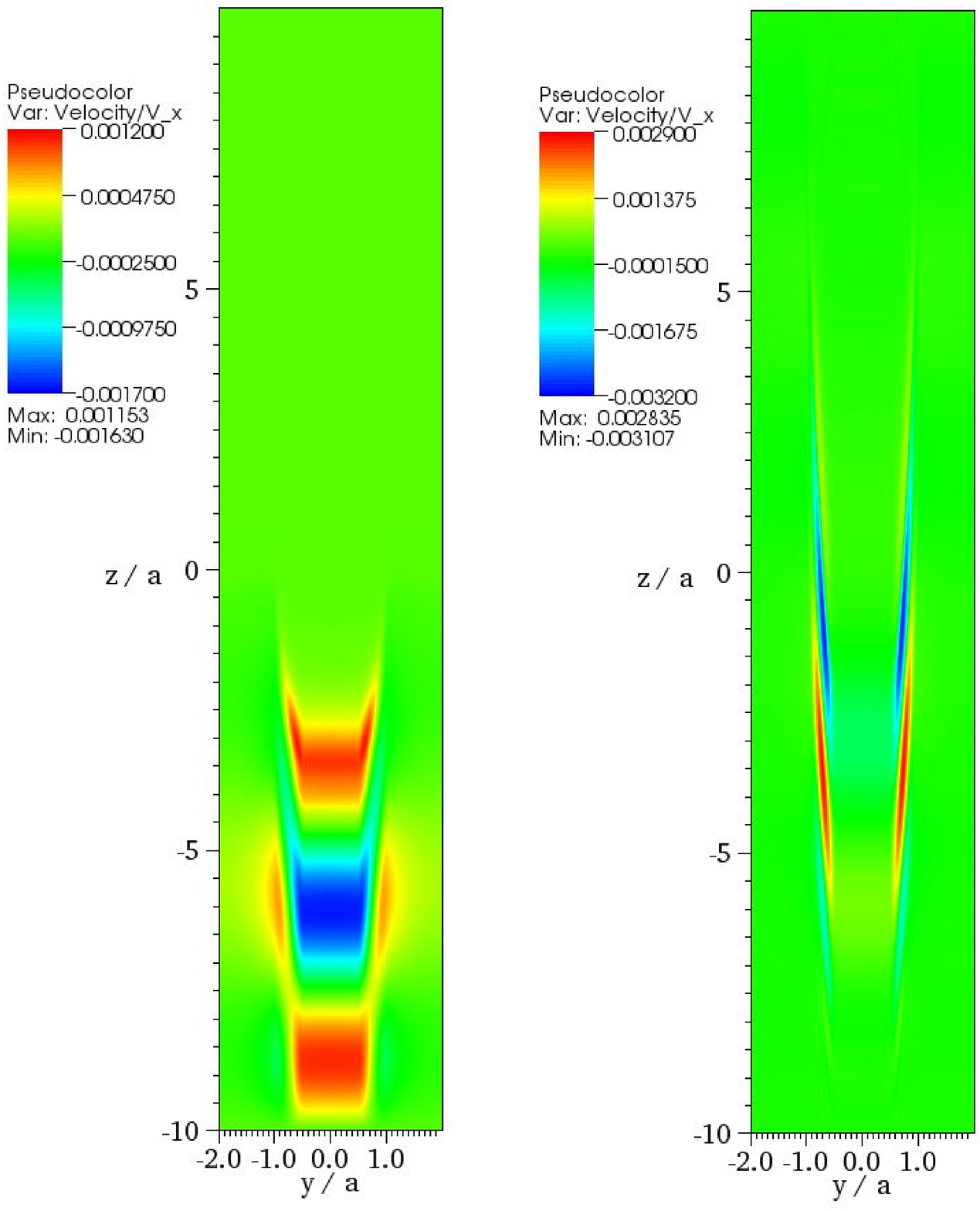}}
\caption{(left) Cartoon illustrating the process of mode coupling for a propagating, transverse footpoint motion, taken from \cite{Pascoe2011}. (right) Snapshots of the transverse velocity (taken from \citealt{2010ApJ...711..990P}) at $t=P$ and $t=5P$ for a density contrast $\rho_0/\rho_e=2$ and a boundary layer $l/a=0.5$.}
\label{fig:modecoupling}
\end{figure}

The interpretation of the propagating transverse waves was clarified in recent 3D, full-MHD numerical simulations \citep{2010ApJ...711..990P,Pascoe2011}. When (even very weak) transverse structuring is present (i.e.~the loop is denser than the surrounding plasma), transverse footpoint motions generate an intrinsic coupling between the kink and (torsional) Alfv\'en modes. This process of mode coupling as the transverse footpoint motion propagates along the loop is similar to the process of resonant absorption in standing modes (see \citealt{Goossens2011} for a review), as illustrated in Fig.\ref{fig:modecoupling} (left);  where the phase speed of the kink mode wave packet matches the local Alfv\'en speed, efficient mode coupling will occur (see e.g.~\citealt{Allan2000}) and energy is transferred rapidly from the transverse motion of the loop (the `kink' mode) to the Alfv\'en mode, concentrated in the shell region of the loop. Hence, the kink mode can be thought of as a moving source of Alfv\'en waves. The (observable) transverse oscillation was found to decay in a few wave periods, which is consistent with observations. Fig.\ref{fig:modecoupling} (right) shows the velocity perturbations after one and 5 periods, respectively. At the early stages of the simulation, just after one full cycle of the footpoint displacement has been completed, the (bulk) transverse oscillation is clearly visible. At later stages, it is clear that the only remaining perturbation is the perturbation in the shell region of the loop, which was identified by \citet{2010ApJ...711..990P} as an ($m=1$) Alfv\'en mode. In a follow-up paper, \cite{Pascoe2011} demonstrate that although some density structuring has to be present to allow the mode coupling to take place, this structuring does not have to be regular (i.e. cylindrically symmetric) and the footpoint motion does not necessarily have to coincide exactly with the density enhancement which forms the loops. 

Several authors have shown that the damping of the transverse motion is frequency-dependent. Focusing on the dominant frequency ($\sim$3.5 mHz) in the COMP observations, \cite{2010ApJ...711..990P} estimate the damping length of the propagating transverse wave packet as 
\begin{equation}
L_d=V_g \tau
\label{eq:Ld}
\end{equation}
where $V_g$ is the group speed of the wave packet and $\tau$ is the damping time as given by 
\begin{equation}
\tau = C {a \over l} {\rho_0 + \rho_e \over \rho_0 - \rho_e} P\,,
\label{eq:RR}
\end{equation}
where $a$ is the loop radius, $l$ the thickness of the shell region, $\rho_0$ and $\rho_e$ are the internal and external densities, respectively and $P$ is the period of the oscillations. The parameter $C$ is a geometric parameter which depends on the specific form of the inhomogeneous layer (see \citealt{Hollweg88,Goossens1992,Ruderman2002}). Combining Eqns.~(\ref{eq:Ld}) and (\ref{eq:RR}), it is clear that 
\begin{equation}
L_d \sim P\,,
\label{eq:TGV}
\end{equation}
i.e.~the longer the period of the footpoint oscillations, the longer the damping length along the loop. Using this expression, \cite{2010ApJ...711..990P}  find that this simple estimate of the damping length is qualitatively consistent with the COMP observation of \cite{Tomczyk2009}. They also show that the mode coupling process is very effective even for modest density contrasts (e.g.~$\rho_0/\rho_e=2$) and quickly tends to an asymptotic value for $\rho_0/\rho_e>4$. This frequency dependent damping is subsequently investigated further by \cite{Terradas2010} and \cite{Verth2010} who show that Eqn.~(\ref{eq:TGV}) effectively leads to a frequency filtering: high-frequency oscillations will be damping faster (i.e.~near the loop footpoints) whereas low-frequency oscillations will be able to propagate further along the loop. \cite{Verth2010} confirm this frequency filtering is present in the data, consolidating the interpretation of the COMP observations as a genuinely coupled kink-Alfv\'en (or `Alfv\'enic') mode. However, it is important to understand that the coupling occurs only in the case of a smooth transverse profile of the plasma structuring, and does not appear when the boundary is a field-aligned discontinuity. Moreover, the {\it observed} (bulk) transverse waves correspond to the kink modes in the central part of the flux tube, while the induced Alfv\'en motions in the tube boundary (the shell region) are currently not resolved. Further recent investigations have focused on additional refinements of the basic model such as partial ionisation (\citealt{Soler2011a}), background flow (\citealt{Soler2011b}) and longitudinal stratification (\citealt{Soler2011c}). Partial ionisation and background flow are found to largely preserve the frequency-dependence in the amplitude decay induced by the mode coupling but longitudinal stratification introduces a more complex picture. Whereas mode coupling causes the amplitude to decrease, longitudinal (gravitational) stratification will cause a competing \emph{increase} in the perturbation amplitudes. As the efficiency of the mode coupling process depends on the frequency, the resulting behaviour of the amplitude will depend on whether the pertubation frequency falls below or above a critical value. This critical value of the frequency depends on the exact parameters of the model, such as the width of the inhomogeneous layer and the density contrast (see Eq.~(52) of \citealt{Soler2011c}).

\section{The Search for (Torsional) Alfv\'en Waves} 
 
In a uniform medium with a straight magnetic field there are four kinds of MHD waves: slow and fast magnetoacoustic waves, the entropy wave and the Alfv\'en wave. The Alfv\'en wave is essentially incompressible, and its group speed is directed strictly parallel with the magnetic field. In non-uniform plasmas, Alfv\'en waves are situated at magnetic flux surfaces (surfaces of constant Alfv\'en speed). In this case, the coordinate along the flux surface plays the role of the ignorable coordinate. The presence of an ignorable, possibly curvilinear, coordinate is necessary, as the wave perturbations must keep the same distance between magnetic field lines. Otherwise, transverse perturbations cannot be incompressible. For example, in a plasma cylinder with a circular cross-section, the ignorable coordinate is the polar angle, and Alfv\'en waves are torsional (i.e.~waves of magnetic twist and plasma rotation). The plasma displacement vector is also locally parallel to the flux surface. This means that Alfv\'en perturbations of neighbouring flux surfaces are disconnected from each other, and they do not constitute a collective mode, in contrast with magnetoacoustic modes of plasma structures. However, often coronal Alfv\'en waves are considered as locally plane, linearly-polarised in the transverse direction that represents the ignorable coordinate. 

If the Alfv\'en speed varies across the magnetic field, Alfv\'en waves situated at different magnetic surfaces experience phase mixing, which leads to the creation of very sharp gradients in the transverse direction. In the presence of small, but finite viscosity or resistivity, these gradients can lead to enhanced dissipation of Alfv\'en waves.  Consider a plane monochromatic Alfv\'en wave with frequency $\omega$, propagating in the $z$ direction in a plasma with a 1D inhomogeneity in the Alfv\'en speed in the $x$ direction, $C_A(x)$. In a developed stage of phase mixing (i.e.~when gradients have developed sufficiently), the wave amplitude decays super-exponentially,
\begin{equation}
V_y(r) \propto \exp \left\{ - \frac{\nu \omega^2}{6 C_A^5(x)}
\left[ \frac{\mathrm{d}C_A(x)}{\mathrm{d}x}\right]^2 z^3\right\},
\label{rrexpz3}
\end{equation} 
where $\nu$ is the shear viscosity \citep{1983A&A...117..220H}. Torsional waves
are also subject to phase mixing, and equation (\ref{rrexpz3}) is applicable to them too.

When considering coronal Alfv\'en waves, especially in open magnetic structures, it is necessary to take into account vertical stratification. A
plane linearly-polarised Alfv\'en wave with wavelength much shorter than the stratification scale height $H$, propagating upwards along a radially-directed magnetic field in an isothermal corona, is governed by the evolutionary equation
\begin{equation}
\frac{\partial V_y}{\partial r} - \frac{R_\odot^2}{4H}
\frac{1}{r^2} V_y - \frac{1}{4C_A(C_A^2-C_s^2)} \frac{\partial
V_y^3}{\partial \tau} -\frac{\nu}{2C_A^3} \frac{\partial^2
V_y}{\partial \tau^2} = 0, 
\label{eveq}
\end{equation}
where $r$ is the vertical coordinate, $R_\odot$ is the radius of the Sun, and $\tau$ is time \citep{2000A&A...353..741N}. The second term describes the change of amplitude with height, the third describes nonlinear effects and the fourth the dissipation. The nonlinearity is connected with the modification of the Alfv\'en speed by the compressible flows induced by the Alfv\'en wave. In the derivation of this equation it was assumed that the nonlinear effects are weak, namely of the same order as the dissipative effects and the ratio of the wavelength to the density scale height. Only the lowest-order nonlinear terms were taken into account. The induced compressible perturbations have double the frequency of the inducing torsional wave. An important feature of the wave evolution in terms of this equation is the singularity at the height where the local Alfv\'en speed is equal to the sound speed. In the case of  propagating weakly-nonlinear long-wavelength torsional waves, this singularity does not appear, since, in a flux tube, the nonlinearly induced compressible perturbation propagates at the sub-Alfv\'enic tube speed \citep{2011A&A...526A..80V}. Moreover, these waves  do not modify the flux tube cross-section, as the centrifugal and magnetic tension forces, associated with the wave perturbations, cancel each other. This difference between torsional and plane Alfv\'en waves should be taken into account in 1D models of coronal Alfv\'en wave dynamics constructed for the study of coronal heating and solar wind acceleration problems (e.g. \citealt{2005ApJ...632L..49S,2010A&A...518A..37M}).

Vertical (field-aligned in open structures) stratification and a vertical change of the magnetic flux tube diameter modify the wave length and hence, affect the efficiency of Alfv\'en wave phase mixing (see, e.g. \citealt{2000A&A...354..334D}). Depending upon the specific geometry, the efficiency of wave damping can either increase or decrease compared to damping in a one-dimensional configuration. In particular, in the case of uniform density, exponentially diverging flux tubes, expression (\ref{rrexpz3}) was found to modify to an $\exp(-\exp(z))$-dependence \citep{1998A&A...338.1118R}.  In addition, the efficiency of phase mixing can be increased by nonlinear steepening (caused by the third term in Eq.~(\ref{eveq}), because of the decrease in wavelength \citep{1998JGR...10323677O}.

Although previous detections of Alfv\'en waves in the solar wind exist (see \cite{Ofman2010} for a comprehensive review or \cite{Gosling2010} for some recent results) and there have been some reports in the recent literature on the possible detection of torsional Alfv\'en waves in the lower solar atmosphere (\citealt{Jess2009}), direct observational detection of torsional waves in the solar corona is still absent. As Alfv\'en waves are essentially incompressible, they cannot be detected with EUV or X-ray coronal imagers, and can be seen only with spectral instruments in those bands. In particular, in a number of studies, unresolved torsional waves were considered as the primary cause of non-thermal broadening of coronal emission lines. The broadening is associated with the Doppler shift caused by unresolved transverse motions of the plasma in the transverse direction. In contrast with kink modes, torsional Alfv\'en waves produce both red and blue Doppler shifts simultaneously, which leads to line broadening. For example, recent analysis of non-thermal broadening measured by Hinode/EIS demonstrated an increase in width of the Fe\,{\sc xii} and Fe\,{\sc xiii} lines in a polar region. The broadening was associated with a non-thermal line-of-sight velocity increase from 26~km s$^{-1}$ at 10'' (i.e.~$\sim$7260 km) above the limb to  42~km s$^{-1}$ some 150'' (i.e.~$\sim$110 000 km) above the limb \citep{2009A&A...501L..15B}. Such behaviour is consistent with a growing Alfv\'en wave amplitude with height, described by the second term in equation (\ref{eveq}). Thus, this analysis can be taken as indirect evidence of torsional waves in the corona. Similar evidence of accelerating propagating disturbances was found by \cite{Gupta2010}, analysing SOHO/SUMER and Hinode/EIS observations of an inter-plume region, who suggested these waves could be either Alfv\'enic or fast magnetoacoustic.

Another possibility for the detection of coronal Alfv\'en waves is open in the microwave band. As the wave changes the local direction of the magnetic field, it can lead to a modulation of the gyrosynchrotron emission (produced by non-thermal electrons accelerated in solar flares), which is sensitive to the angle between the magnetic field in the emitting plasma and the line-of-sight. Hence, in the presence of a torsional wave, the microwave emission can be periodically modulated. Evidence of this process was found by \cite{2003ApJ...588.1163G} with the use of the Nobeyama Radioheliograph. However, the search for coronal Alfv\'en waves in the microwave band requires more attention.

\section{Quasi-Periodic Pulsations} 
 
Oscillatory variations or quasi-periodic pulsations (QPP) of radio emission generated in solar flares have been investigated for several decades (see, e.g., \citealt{1987SoPh..111..113A}).  The periodicities range from a fraction of a second to several minutes, and the modulation depth of the emission reaches 100\%.  Similar periodicities are often seen in hard X-rays (see \cite{2009SSRv..149..119N} for a review) and were recently found in gamma-rays \citep{2010ApJ...708L..47N}. There is growing evidence that QPP are a common and perhaps even intrinsic feature of solar flares: recent analysis of microwave emission generated in twelve similar single-loop flares showed that ten events (83\%) had at least one or more significant spectral component with periods from 5--60 s \citep{2010SoPh..267..329K}.  Likewise, well-pronounced QPP have been found in radio, white-light and soft X-ray lightcurves  of stellar flares \citep{2003A&A...403.1101M, 2005A&A...436.1041M}. However, no systematic study of this phenomenon has been carried out. 
 
In the context of MHD coronal seismology, flaring QPP are a very interesting subject. Firstly, the observed periods of QPP coincide with the periods of coronal waves and oscillations that are confidently interpreted in terms of MHD wave theory. Thus, QPP may well be the manifestation of coronal MHD oscillations, either modulating the emission directly, by changing the density, temperature and the absolute value or the direction of the magnetic field in the emitting plasma, or affecting the dynamics of the charged non-thermal particles (e.g., \citealt{2008PhyU...51.1123Z}), or periodically triggering magnetic reconnection and/or affecting the efficiency of the charged particle acceleration (\citealt{2005A&A...440L..59F, 2006SoPh..238..313C}). A possible mechanism is the generation of anomalous resistivity by current-driven instabilities in current density spikes produced by magnetoacoustic waves in the vicinity of reconnection sites, which triggers magnetic reconnection. 
 
Strong evidence supporting the connection of QPP with MHD oscillations is the observation of multiple periodicities, which may be associated with different spatial harmonics of MHD modes of oscillating structures. For example, the simultaneous presence of 28 s, 18 s, and 12 s oscillations was found in the microwave and hard X-ray emission in a flare \citep{2009A&A...493..259I}. The ratio of the periods suggests that the oscillations are likely to be produced by different spatial harmonics of the highly-dispersive sausage mode.  
 
Thus, the observation of QPP provides us with a powerful tool for the detection of MHD waves in coronal plasma structures and hence for MHD coronal seismology. The study of MHD oscillations in flaring emission has several advantages in the context of coronal seismology. In particular, as flares are the brightest events in solar physics, their observations can be of much higher cadence than the observations of the quiet corona. This allows one to achieve sub-second time resolution, which is, in particular, necessary for the detection of the fast-wave travel time across a coronal plasma structure. The latter time scale is needed for the diagnostics of fine transverse structuring of the corona (see, e.g., \citealt{2005SSRv..121..115N}). Moreover, flaring energy releases are efficient drivers of MHD oscillations in the plasma structures around the flaring site, as the solar corona is a highly elastic and compressible medium. These oscillations, in turn, can modulate the flaring emission.  This modulation can also be produced by MHD modes which have not been detected in the corona by other methods. For example the essentially incompressible torsional (or Alfv\'en) waves can modulate the efficiency of the gyrosynchrotron emission by changing the direction of the magnetic field to the line-of-sight \citep{2003ApJ...588.1163G}.  Because of the strong dependence of the emission on the line-of-sight angle, the relative amplitude of the observed modulation can be several times higher than the amplitude of the wave, maximising the chances for its detection. 
 
QPP can be produced not only by MHD oscillations, but may also directly result from oscillatory regimes of energy releases, e.g. periodic shedding of plasmoids \citep{2000A&A...360..715K} or over-stability of a current sheet with an externally generated steady flow \citep{2006ApJ...644L.149O}. Often oscillatory or quasi-oscillatory energy releases are observed in massive numerical simulations of MHD processes associated with solar flares (e.g. in experiments on magnetic flux emergence \citep{2009A&A...494..329M}), where it is difficult to determine the cause of the oscillatory behaviour. This group of mechanisms for QPP is often referred to as a \lq\lq dripping model" or \lq\lq load/unload" \citep{2010PPCF...52l4009N}, as essentially it is based upon the conversion of a steady energy supply (e.g. the reconnecting plasma inflows) into a periodic energy release. Mathematically, it can be considered in terms of auto-oscillations of the dynamical system, such as limit cycles and relaxation oscillations. As the observed parameters (periods, amplitudes, signatures) of auto-oscillations are independent of the initial excitation and are determined by the parameters of the system only (in the case of periodic magnetic reconnection it could be the inflow rate, plasma resistivity and density, and the strength of the guiding field), their identification has very promising seismological potential which remains to be explored. One of the expected benefits is revealing the physical processes responsible for the magnetic energy release in flares, as the identification of the mechanisms for QPP puts additional constraints on the models of solar flares. Independently of the specific mechanism for the generation of QPP, their study opens up unique opportunities for seismology of stellar coronae. 
 
A major recent avenue in the study of QPP is the investigation of their spatial structure with the use of X-ray imagers and microwave interferometers. Similarly to the morphology of the flares, QPP can appear either in a single loop geometry, when the hard X-ray sources are situated at the footpoints and sometimes at the loop apex (e.g. \citealt{2008A&A...487.1147I}), while the microwave QPP occur in the legs of the loop \citep{2010SoPh..267..329K}. Hard X-ray and microwaves QPP are usually synchronous and hence are very likely produced by the same population of non-thermal electrons, accelerated in the energy release.  
 
Another class of QPP is observed in two-ribbon flares, where the individual  
bursts of the emission come from different loops situated along a neutral line in the flaring arcade (see, e.g., \citealt{2009SoPh..258...69Z}). The speed of the energy release progression  along the neutral line is typically a few tens km/s, which is at least an order of magnitude lower than the sound and Alfv\'en speeds. Very recently, this lower speed was shown to be consistent with the group speed of slow magnetoacoustic waves guided by an arcade \citep{2011ApJ...730L..27N}. In that model, a weakly oblique slow magnetoacoustic wave bounces between the arcade's footpoints and gradually progresses across the field. The highest value of the group speed occurs at a propagation angle of 25$^\circ$--28$^\circ$ to the magnetic field. This effect can explain the temporal and spatial structure of quasi-periodic pulsations observed in two-ribbon flares: the sites of the energy releases gradually move along the axis of the arcade, across the magnetic field, at the group speed of the slow magnetoacoustic waves. The time between the individual bursts can be estimated as the slow wave travel time from the apex of the arcade to the footpoints and back, which is also consistent with observations. Another interesting topic is the generation of waves by flares. Parameters of these wave, in particular the periods, are determined by the parameters of the flares (see, e.g.,\citealt{Liu2011}).
  
\section{Discussion \& Open Questions}

After an initial period of very rapid growth following the launch of SOHO and TRACE in the late 1990's, coronal seismology is now going through a period of consolidating results and developing theoretical models to include more detailed structuring and additional physical processes, as well as incorporating new observational results from e.g.~Hinode and SDO. The coronal flux tube model (\citealt{Zaitsev1975,Edwin1983}) still forms the basis of much of the modelling and the interpretation of observed waves and oscillations. However, it is becoming evident that the simplified linear model must be extended considerably, taking into account realistic geometry, nonlinearity, and dissipation, so that coronal seismology can become a more accurate diagnostic tool. Again, a detailed description of all recent results is beyond the scope of this brief review so we highlight a few areas that have recently been reported in the literature.

Various aspect of both transverse and longitudinal fine-structuring have been investigated. The period-ratio $P_1/2P_2$ has received a lot of attention, as its deviation from unity has implications for the longitudinal structuring of coronal loops and hence has potential as a seismological tool (see \cite{Andries2009} for a review and \cite{Macnamara2010,Macnamara2011} for more recent results). Whereas the modelling of the period ratio focuses on the effect of loop structuring on the temporal evolution of the wave modes, the spatial deformation of the mode and its seismological implications have also been studied (\citealt{Erdelyi07,VerthErdelyi08,Pascoe2009,Verth07,Verth08,Verth2011}). Most coronal loops are likely to be cooling as they are observed to oscillate, an aspect which had not previously been included in the modelling. A series of papers has recently studied how this cooling affects the damping rate of the oscillations (\citealt{Morton2009,Morton2010,Mortonetal2010,Ruderman2011}). Furthermore, numerical studies have also started to focus on genuine 3D modelling of coronal loop oscillations (e.g.~\citealt{Pascoe2009b,Selwa2011a,Selwa2011b,Ofman2009} for a review) and on the effect of the magnetic field structure, i.e.~the magnetic topology, on the behaviour of MHD waves (e.g.~\citealt{McLaughlin2009,Selwa2010} and \cite{McLaughlin2011} for a review). Finally, we point out that efforts to refine the estimates of the coronal magnetic field and improve their accuracy are ongoing (\citealt{Erdelyi08,IDM-Pascoe09,Verwichte2009,Terradas2011}) and that a first attempt was made to estimate the adiabatic index using Hinode/EIS observations \citep{VanDoorsselaere2011}.\\

The topics we highlight in this review were chosen because they are currently receiving substantial attention in the literature, either as newly emerging observations of waves and oscillations or because the established interpretation is being challenged. Rather than providing a further summary, we conclude this review with a series of open questions for each of the four highlighted topics.
 
\subsection{Quasi-periodic Propagating Disturbances}

- What is the root cause of the periodic disturbances (i.e.~how are they generated) and what determines their observed speeds and periodicities?\\
- What are the observational signatures of waves and flows in the outer solar atmosphere; will we ever be able to discriminate or are both waves and flows present?\\
- If the propagating disturbances are flows, how can they propagate with a speed equal to the sound speed without forming shocks? If they are waves, is there clear evidence that the propagation speed corresponds to the local sound speed at all temperatures? \\
- Why are the propagating disturbances only seen for short distances along coronal structures? If they are waves, what is the relevant damping mechanism? If they are flows, what happens to the material?\\
- What role do these quasi-periodic propagating disturbances play in heating and/or mass loading of the outer solar atmosphere?

The waves and flows interpretations can each explain some of the observed properties, but neither can currently account for all of the observational signatures. As the observational data have been pushed to their (current) limits, progress might have to be made by studying realistic numerical models, combined with forward modelling to identify observable signatures for either, or both, interpretations.

\subsection{Propagating Transverse Loop Oscillations}

- Are these waves truly ubiquitous, as they appear to be from recent observations and how are they generated?\\
- Can the Alfv\'en waves induced by the decaying kink waves through mode coupling in the loop boundary be observed?\\
- Can they really provide enough energy to heat the (quiet) solar corona and/or drive the solar wind? \\
- How can they be used as a seismological tool - i.e.~what can we learn about the coronal plasma or loops from these observations? Current models strongly indicate the presence of unresolved transverse structuring and hence we should be able to use the observed propagating transverse waves to probe the unresolved fine structure.

\subsection{Torsional Alfv\'en Waves}

- Can we observe them directly with current or future missions (e.g.~Solar Probe) or will we only be able to achieve indirect observations in the solar corona? \\
- Is the MHD description sufficient, or is the localisation of the theoretically predicted phase-mixed Alfv\'en perturbations more adequately described at the kinetic level?\\
- Can the waves tell us something about the field-aligned electric currents and hence be used to assess the relevance and precision of non-linear force-free magnetic field extrapolations and the amount of free magnetic energy available in active regions?

 \subsection{Quasi-periodic Pulsations (QPP)}

- What are they and how are they generated?\\
- Does the auto-oscillatory regime of magnetic reconnection really exist? \\
- What can QPP tell us about the physical conditions and processes operating in reconnection sites?\\
- Are QPP connected with MHD oscillations and waves in reconnection sites and what is the nature of this relationship? Can MHD scale waves affect local (dissipation scale) processes?\\

With the advent of SDO, providing high-resolution, full disk observations in a range of wavelengths, the number of observed waves and oscillations is bound to increase rapidly in the near future. This will allow coronal seismology to start making statistical studies to confirm the properties induced so far from a relatively small number of observations. At the same time, it will be crucial to combine this information gained from the imaging telescopes with spectroscopic observations, providing information about the local plasma, and to use the entire spectrum of wavelengths (e.g.~radio) currently available through a combination of ground and space-based instrumentation. Finally, a two-pronged approach is needed on the modelling side; further refinement and in-depth study of the simple but informative basic models, combined with numerical simulations of realistic coronal (active region) configurations.

\begin{acknowledgements} 
The authors would like to thank S.W.~McIntosh for providing Fig.~\ref{fig:Tomczyk}. IDM acknowledges support from a Royal Society University Research Fellowship. 
\end{acknowledgements} 
 
\bibliographystyle{rspublicnat}

\bibliography{references}

\end{document}